\input harvmac
\input epsf.tex
\noblackbox
\overfullrule=0pt
\def\Title#1#2{\rightline{#1}\ifx\answ\bigans\nopagenumbers\pageno0\vskip1in
\else\pageno1\vskip.8in\fi \centerline{\titlefont #2}\vskip .5in}

%
%
%

\def\[{\left [}
\def\]{\right ]}
\def\({\left (}
\def\){\right )}

\def\p{\partial}

\def\RN{Reissner-Nordstr\"om}
\font\cmss=cmss10 \font\cmsss=cmss10 at 7pt
\def\IZ{\relax\ifmmode\mathchoice
   {\hbox{\cmss Z\kern-.4em Z}}{\hbox{\cmss Z\kern-.4em Z}}
   {\lower.9pt\hbox{\cmsss Z\kern-.4em Z}}
   {\lower1.2pt\hbox{\cmsss Z\kern-.4em Z}}\else{\cmss Z\kern-.4emZ}\fi}

%
\lref\bars{ I. Bars, Phys. Rev. {\bf D55} (1997) 2373, hep-th/9607112.}

\lref\vafaf{ C. Vafa, Nucl. Phys. {\bf B469} (1996) 403, hep-th/9602022.}
\lref\swsix{ E. Witten, hep-th/9507121, N. Seiberg and E. Witten,  Nucl. Phys. {\bf
B 471 } (1996) 121, hep-th/9603003}
\lref\abkss{ O. Aharony, M. Berkooz, S. Kachru, N. Seiberg and E. Silverstein,
hep-th/9707079.}

\lref\hawellis{S. Hawking and J. Ellis, {\it The  large scale structure 
of spacetime}, Cambrige Univ. Press (1973), and references therein. }

\lref\dr{M. Dine and A. Rajaraman,  hep-th/9710174. }

\lref\dos{ M. Douglas, H. Ooguri and S. Shenker,   Phys. Lett. {\bf B402} (1997) 36, hep-th/9702203; 
O. Ganor, R. Gopakumar and S. Rangoolam, hep-th/9705188;
M. Douglas and H. Ooguri,  hep-th/9710178. }

\lref\andyfive{
A. Srominger,  Phys. Lett. {\bf B383} (1996) 44, hep-th/9512059.
}

\lref\hlm{G. Horowitz, D. Lowe and J. Maldacena, 
Phys. Rev. Lett. {\bf 77} (1996) 430, hep-th/9603195.}

\lref\hms{G. Horowitz, J. Maldacena and A. Strominger,
 Phys. Lett. {\bf B383} (1996) 151,
 hep-th/9603109.}
\lref\dbr{J. Polchinski, S. Chaudhuri, and C. Johnson, hep-th/9602052.}

\lref\ghas{G. Horowitz and A. Strominger,
Phys. Rev. Lett. {\bf 77} (1996) 2368,
  hep-th/9602051.}
\lref\ascv{A. Strominger and C. Vafa,
 Phys. Lett. {\bf B379} (1996)
99,
 hep-th/9601029.}

\lref\spin{
 J.C. Breckenridge, R.C. Myers, A.W. Peet and  C. Vafa, 
Phys. Lett. {\bf B391} (1997) 93, hep-th/9602065.}

\lref\cama{C. Callan and J. Maldacena,
 Nucl. Phys. {\bf B 475 } (1996)
645, hep-th/9602043.}


\lref\jmas{J. Maldacena and A. Strominger, 
hep-th/9609026.}

\lref\bfss{
 T. Banks, W. Fischler, S.  Shenker and  L. Susskind,
hep-th/9610043.}

\lref\dps{ 
M. Douglas, J. Polchinski and A. Strominger, 
hep-th/9703031.}

\lref\lennyfiniten{
L. Susskind, 
  hep-th/9704080.}

\lref\hstro{ G. Horowitz and A. Strominger,
Nucl. Phys. {\bf B360} (1991) 197. }




\lref\msfive{ J. Maldacena and A. Strominger, hep-th/9710014.}

\lref\nahm{W.  Nahm, Nucl. Phys. {\bf B135} (1978) 149.}

\lref\kt{ I. Klebanov and A. Tseytlin, Nucl. Phys. {\bf B475} (1996) 
164,
 hep-th/9604089.}

\lref\jmstrings{ J. Maldacena,  Proceedings of Strings'97, hep-th/9709099. }

\lref\seimm{N. Seiberg, Phys. Rev. Lett. {\bf 79} (1997) 3577,
 hep-th/9710009.}

\lref\senmm{ A. Sen, hep-th/9709220.}

\lref\bankssei{ T. Banks and N. Seiberg, Nucl. Phys. {\bf B497} (1997) 41,
hep-th/9702187.}

\lref\seiberg{ N. Seiberg,  hep-th/9705117.}

\lref\jmmlow{ J. Maldacena, Phys. Rev. {\bf D55} (1997) 7645,
hep-th/9611125.}
\lref\edhiggs{E. Witten, hep-th/9707093.}

\lref\seidia{ D. Diaconescu and N. Seiberg,  JHEP07(1997)001, hep-th/9707158.}

\lref\koreano{ S. Hyun, hep-th/9704005.}

\lref\kostas{ H. Boonstra, B. Peeters and  K. Skenderis,
 hep-th/9706192}
\lref\sfkostas{  K. Sfetsos and  K. Skenderis, hep-th/9711138.}

\lref\lsss{ S. Sethi and L. Susskind, 
Phys. Lett. {\bf B400} (1997) 265, hep-th/9702101.}

\lref\polch{S. Hellerman and  J. Polchinski,
hep-th/9711037.}

\lref\greycalc{ A. Dhar, G. Mandal and  S. Wadia
Phys. Lett. {\bf B388} (1996) 51, hep-th/9605234; 
S. Das and S. Mathur, 
Nucl. Phys. {\bf B478} (1996) 561, hep-th/9606185; Nucl.Phys. {\bf B482}
 (1996) 153,
hep-th/9607149; J. Maldacena and A. Strominger,
Phys. Rev. {\bf D55} (1997) 861, 
hep-th/9609026;
 S. Gubser, I. Klebanov,
 Nucl. Phys. {\bf B482} (1996) 173,
 hep-th/9608108;  C. Callan, Jr., S. Gubser, I. Klebanov and A. Tseytlin,
 Nucl. Phys. {\bf B489} (1997) 65,
hep-th/9610172;
 I. Klebanov and  M. Krasnitz
Phys. Rev. {\bf D55} (1997) 3250,
hep-th/9612051;
I. Klebanov, A. Rajaraman and  A. Tseytlin 
Nucl. Phys. {\bf B503} (1997) 157,
 hep-th/9704112; S. Gubser, hep-th/9706100; H. Lee, Y. Myung and
J. Kim, hep-th/9708099;  K. Hosomichi, hep-th/9711072.}

\lref\finn{ M. Cvetic and F. Larsen,  Phys. Rev. {\bf D56} (1997) 4994,
 hep-th/9705192; hep-th/9706071. }

\lref\msw{ J. Maldacena, A. Strominger and E. Witten, hep-th/9711053. }

\lref\msangular{ J. Maldacena and A. Strominger, Phys. Rev. {\bf D56}
(1997) 4975,
 hep-th/9702015.}

\lref\seiprivate{O. Aharony, S. Kachru,  N. Seiberg and E. Silverstein, private communication.}

\lref\polstring{A. Polyakov, hep-th/9711002.}

\lref\bfks{ T. Banks, W. Fishler, I. Klebanov and L. Susskind, 
hep-th/9709091.}

\lref\entrothree{ S. Gubser, I. Klebanov and A. Peet, 
Phys. Rev. {\bf D54} (1996) 3915,  hep-th/9602135;  A. Strominger, 
unpublised. }

\lref\dtgrey{\bf do away with this}

\lref\bfsse{ T. Banks, W. Fishler, N. Seiberg and L. Susskind, 
Phys. Lett. {\bf B408} (1997) 111,
 hep-th/9705190.}

\lref\schwarz{ 
M. Aganagic, C. Popescu and  J. Schwarz,
Nucl. Phys. {\bf B495} (1997) 99,
hep-th/9612080. }
 
\lref\dinesei{M. Dine and N. Seiberg, 
 Phys. Lett. {\bf B409} (1997) 239,  hep-th/9705057.}

\lref\kledtdil{ I. Klebanov,  Nucl. Phys. {\bf B496} (1997) 231,
 hep-th/9702076.}

\lref\dtnren{ S. Gubser and I. Klebanov, hep-th/9708005.}

\lref\dtgrav{S. Gubser,  I. Klebanov and A. Tseytlin, 
Nucl. Phys. {\bf B499} (1997) 217,
 hep-th/9703040.}

\lref\btz{ Ba{\~n}ados, Teitelboim and  Zanelli, 
Phys. Rev. Lett. {\bf 69}
(1992) 1849, hep-th/9204099.}

\lref\irish{ D. Birmingham,  I. Sachs and  S. Sen,
hep-th/9707188.}
 
\lref\siens{ N. Seiberg, Phys. Lett. {\bf B408} (1997) 98, 
 hep-th/9705221 }

\lref\carlip{ S. Carlip, Phys. Rev. {\bf D51} (1995) 632, gr-qc/9409052. }

\lref\bbpt{  K. Becker, M. Becker,
J. Polchinski and 
 A. Tseytlin,
Phys. Rev. {\bf D56} (1997) 3174,
 hep-th/9706072. } 

\lref\fixedmod{ 
S. Ferrara, R. Kallosh and A. Strominger, Phys. Rev. {\bf D52} (1995) 5412,
hep-th/9508072; S. Ferrara and R. Kallosh, Phys. Rev. {\bf D54} (1996) 1514, 
hep-th/9602136; Phys.Rev. {\bf D54} (1996) 1525, hep-th/960309.
}

\lref\ferrarafm{L. Andrianopoli, R. D'Auria and
 S. Ferrara,  Int. J. Mod .Phys {\bf A12} (1997) 3759,  hep-th/9612105.}

\lref\fmf{ A. Chamseddine, S. Ferrara, G. Gibbons and R Kallosh,
Phys. Rev. {\bf D55} (1997) 3647, hep-th/9610155.}

\lref\enhanc{ 
G.  Gibbons, Nucl. Phys. {\bf B207}, (1982) 337;
R. Kallosh and A. Peet, Phys. Rev. {\bf D46} (1992) 5223, hep-th/9209116;
 S. Ferrara, G. Gibbons, R.
Kallosh, Nucl. Phys. {\bf B500} (1997) 75, hep-th/9702103.}

\lref\ggpt{
G. Gibbons and P. Townsend, Phys. Rev. Lett. {\bf 71} (1993) 5223,
hep-th/9307049.}

\lref\sohnius{
R. Haag, J. Lopuszanski and M. Sohnius, Nucl. Phys. {\bf B88} (1975) 
257.}

\lref\ckvp{P. Claus, R. Kallosh and A. van Proeyen, 
hep-th/9711161.}

\lref\bduff{ M. Blencowe and M. Duff, Phys. Lett. {\bf 203B} (1988)
229; Nucl. Phys. {\bf B310} (1988) 387.}

\lref\nst{ H. Nicolai, E. Sezgin and Y. Tanii, Nucl. Phys. {\bf B305}
(1988) 483. }

\lref\dgt{ M. Duff, G. Gibbons and P. Townsend, 
hep-th/9405124.}

\lref\ght{G. Gibbons, G. Horowitz and P. Townsend, 
hep-th/9410073.}

\lref\gun{
M. G\"unaydin and N. Marcus, Class. Quant. Grav. {\bf 2} (1985) L11;
 Class. Quant. Grav. {\bf 2} (1985) L19;
H. Kim, L. Romans and P. van Nieuwenhuizen, Phys. Lett. {\bf 143B} (1984)
 103;
M. G\"unaydin, L. Romans and N. Warner, Phys. Lett. {154B} 
(1985) 268; Phys. Lett. {164B} 
(1985) 309; Nucl. Phys. {\bf B272} (1986) 598
}

\lref\sugradiv{
Look for ``gauged'' supergravities in 
{\it Supergravities in Diverse Dimensions }, Vol. 1 and 2, 
                           A. Salam and  E. Sezgin, (1989), 
North-Holland. 
}

\lref\singleton{
C. Frondal, Phys. Rev. {\bf D26} (82) 1988;
D. Freedman and H. Nicolai, Nucl. Phys. {\bf B237} (84) 342;
K. Pilch, P. van Nieuwenhuizen and P. Townsend, Nucl. Phys. {\bf B242}
(84) 377;
M. G\"unaydin, P. van Nieuwenhuizen and N. Warner, 
Nucl. Phys. {\bf B255} (85) 63; 
M. G\"unaydin and N. Warner, Nucl. Phys. {\bf B272} (86) 99;
M. G\"unaydin, N. Nilsson, G. Sierra and P. Townsend, Phys. Lett. {\bf
 B176} (86) 45; 
 E. Bergshoeff, A. Salam, E. Sezgin and Y. Tanii, Phys. Lett. {\bf
205B} (1988) 237; Nucl. Phys. {\bf D305} (1988) 496; 
E. Bergshoeff, M. Duff, C. Pope and E. Sezgin, Phys. Lett. {\bf B224} (1989)
71; 
}

\lref\sltwo{L. Susskind, hep-th/9611164;
O. Ganor, S. Ramgoolam and  W. Taylor IV, Nucl. Phys. {\bf B492} 
(1997) 191, hep-th/9611202.
}

\lref\mss{J. Maldacena and L. Susskind, 
Nucl .Phys. {\bf B475} (1996) 679, hep-th/9604042.}

\lref\kal{
 R. Kallosh, J. Kumar and  A. Rajaraman, hep-th/9712073.
}

    
%
\Title{\vbox{\baselineskip12pt
\hbox{hep-th/9711200}\hbox{ HUTP-97/A097}}}
{\vbox{
{\centerline { The Large N Limit of Superconformal field }}
{\centerline {theories and supergravity }}
  }}
\centerline{Juan Maldacena\foot{malda@pauli.harvard.edu}}
\vskip.1in
\centerline{\it Lyman Laboratory of Physics, Harvard University,
Cambridge, MA 02138, USA}
\vskip.1in
\vskip .5in

\centerline{\bf Abstract}

We show that  the large $N$ limit of certain  conformal field theories 
in various dimensions include in their Hilbert space 
a sector describing supergravity 
on the product of Anti-deSitter spacetimes,
spheres and other compact manifolds. 
This is shown by taking some branes in the full
M/string theory and then taking a low energy limit 
where the field theory
on the brane decouples from the bulk.
We  observe that, in this limit, we can
still  trust the near horizon geometry for large $N$. 
The enhanced supersymmetries of the near horizon geometry correspond
to the extra supersymmetry generators present in the superconformal
group (as opposed to just the super-Poincare group).
The 't Hooft limit of 3+1  ${\cal N} =4$ super-Yang-Mills 
at the conformal point is shown 
to contain strings: they are IIB strings. 
We conjecture that compactifications  of M/string theory on 
various Anti-deSitter spacetimes is dual to various conformal
field theories. This leads to   a new proposal for  a  definition of M-theory
which could be extended to include 
five  non-compact dimensions.

 \Date{}

%

\newsec{General idea }

In the last few years it has been extremely fruitful to derive 
quantum field theories by taking various limits of  string or M-theory.
In some cases this is done by considering the theory  at 
geometric singularities 
and in others by considering a configuration containing branes 
and then taking a limit where the dynamics on the brane decouples 
from the bulk. 
In this paper we  consider theories that are obtained by 
decoupling  theories on  branes from gravity. 
We  focus on 
conformal invariant field theories but a similar analysis 
could be done for non-conformal field theories.
 The cases considered include 
$N$ parallel D3 branes in IIB string theory and various others.
We  take the limit where the field theory on the brane  decouples from the 
bulk. At the same time we look at the near horizon geometry 
and we argue that the supergravity solution can be trusted as long
as $N$ is large. $N$ is kept  fixed as we take the 
 limit.
The approach is similar to that used in \msfive\ to study the
NS fivebrane theory \siens\ at finite 
temperature.
The supergravity solution typically reduces to $p+2$ dimensional
Anti-deSitter space 
($AdS_{p+2}$) 
times  spheres (for D3 branes we have  $AdS_5\times S^5$).
The curvature  of the sphere and the 
$AdS$ space in Planck units
is a (positive) power of $1/N$. 
Therefore the solutions can be trusted as long as $N$ is large. 
Finite temperature configurations  in the decoupled field
theory correspond to black hole configurations in $AdS$ spacetimes.
These black holes will Hawking radiate into the $AdS$ spacetime.
We conclude that excitations of the $AdS$ spacetime 
are included in  the Hilbert space of the corresponding
conformal field theories. 
A theory in $AdS$ spacetime is not completely well defined since
there is a horizon and it is also necessary to give 
some boundary conditions at infinity. 
 However, local properties
and local processes can be calculated  in supergravity when
 $N$  is large if the proper energies
involved are much bigger than the energy scale set by the cosmological
constant (and smaller than the Planck scale).
We will conjecture that the full quantum M/string-theory on $AdS$ space,
 plus suitable 
boundary conditions is dual to  the  corresponding brane theory.
We are not going to specify the boundary conditions in $AdS$,
  we leave this interesting problem 
for the future. The $AdS \times$(spheres) description will become
useful  for large $N$, where we can isolate some local processes from
the question of  boundary conditions.
The supersymmetries of both theories agree, both are given by the
superconformal group.
The superconformal group has twice the
amount of supersymmetries of  the  corresponding 
super-Poincare group\refs{\sohnius,\nahm}. 
This enhancement of supersymmetry near the horizon
of extremal black holes 
 was observed in \refs{\enhanc,\ggpt} precisely by showing that
the near throat geometry reduces to $AdS \times$(spheres). 
$AdS$ spaces (and branes in  them) were extensively considered in
the  literature 
\refs{\sugradiv,\singleton,\gun,\bduff,\nst,\dgt,\ckvp},
includding the   connection with the superconformal group.

In section 2 we study ${\cal N} =4$  d=4 $U(N)$
super-Yang-Mills as a first example, we discuss several issues which
 are present in all  other cases.
In section 3
we analyze  the theories describing M-theory five-branes and M-theory
two-branes.
 In section 4
 we consider theories with  lower supersymmetry which are related
to a 
 black string in six dimensions  made with D1 and D5 branes.
In section 5 we study theories with even less supersymmetry
involving 
black strings 
in five dimensions and finally we mention 
the theories related to extremal \RN\ black holes in four  
spacetime dimensions (these last cases will be more speculative and contain
some unresolved puzzles). 
Finally in section 6 we make some comments on the relation to 
matrix theory.

\newsec{ D3 branes or  ${\cal N} =4$  $U(N)$ super-Yang-Mills in d=3+1}

We start with type IIB string theory with string coupling $g$, which will 
remain fixed. Consider $N$ parallel D3 branes separated by some distances
which  we denote by $r$.
For low energies the theory on the D3 brane decouples from the bulk.
It is more convenient to take the energies fixed and take
\eqn\decoupling{
 \alpha' \to 0 ~,~~~~~~~~~~~~~ 
U \equiv {r \over \alpha'} = {\rm fixed}~. 
}
The second condition is saying that we keep 
 the mass of the stretched strings fixed.
 As we take the decoupling limit we bring the branes together 
but the  the Higgs expectation values  corresponding  to this separation
remains fixed. The resulting  theory on the brane is 
 four dimensional  ${\cal N} =4$  $U(N)$ SYM. 
Let us  consider the theory at the superconformal point, where $r=0$.
The conformal group is SO(2,4). We also have an $SO(6)\sim SU(4)$
 R-symmetry that
rotates the six scalar fields  into each  other\foot{
The representation includes objects in the  spinor representations, so
we should be talking about SU(4), we will not make this, or similar 
distinctions in what follows.}.
The superconformal group includes twice the number of supersymmetries 
of the super-Poincare group: the commutator of  special conformal
transformations with  Poincare supersymmetry generators gives the new 
supersymmetry generators. The precise superconformal algebra was
computed in \sohnius  .  All this is valid for any $N$.

Now we consider the supergravity solution carrying D3 brane charge 
 \hstro
\eqn\dthree{\eqalign{
ds^2 &= f^{-1/2} dx_{||}^2  + f^{1/2} (dr^2 + r^2 d\Omega^2_5 )~,
 \cr
f & = 1 + { 4 \pi  g N \alpha'^2 \over r^4 }~,
}}
where $x_{||}$ denotes the four coordinates along the worldvolume of
the three-brane and $d\Omega^2_5$ is the metric on the unit five-sphere\foot{
We choose conventions where $g\to 1/g $ under S-duality.}.
The self dual five-form field strength is nonzero and has a flux on the 
five-sphere.  
Now we define the new variable $U \equiv {r \over \alpha'}$ and
we rewrite the metric in terms of $U$. Then we take the $\alpha' \to 0$
limit. Notice that  $U$ remains fixed.
In this  limit we can neglect the
1 in the harmonic function \dthree .
The metric becomes 
\eqn\nh{
ds^2 = \alpha' \left[ { U^2 \over \sqrt{ 4\pi g N }} dx_{||}^2 
   + \sqrt{ 4\pi g  N  } {dU^2 \over U^2} + 
\sqrt{  4\pi g N  } d \Omega_5^2  \right] ~.
}
This metric describes  five dimensional Anti-deSitter 
 ($AdS_5$) times 
a five-sphere\foot{
See the appendix for a brief description of $AdS$ spacetimes.}.
We see that there is an overall $\alpha'$ factor. 
The metric remains constant in $\alpha'$ units. The radius
of the five-sphere is $R^2_{sph}/\alpha' = \sqrt{4\pi g N} $, and  is the 
same as the ``radius'' of $AdS_5$ (as defined in the appendix). 
In ten dimensional Planck units they are both proportional to $N^{1/4}$.
The radius is quantized because the flux of the 5-form field strength on the
5 sphere is quantized.  We can trust the supergravity solution when
\eqn\sugra{
   gN \gg 1 ~.
}
When  $N$ is  large  we have  approximately 
ten dimensional  flat space in the
  neighborhood of any  point\foot{
In writing \sugra\ we assumed
that $g\le 1$, if $g>1$ then the condition is $ N/g \gg1 $.  In other words
we need large $N$ , not large $g$.}. 
Note that in the large $N$ limit the flux of the 5 form field strength
per unit Planck (or string) 5-volume becomes small.

Now consider a near extremal black D3 brane solution 
in the decoupling limit \decoupling . We  keep
the energy density  
on the brane worldvolume theory ($\mu$) fixed. 
We find the metric
\eqn\ne{\eqalign{
ds^2 =& \alpha' \left\{ { U^2 \over \sqrt{4 \pi g N }} 
\left[ - (1- U_0^4/U^4) dt^2 
+ dx_i^2 \right] 
   + \sqrt{ 4 \pi g N } {dU^2 \over U^2(1- U_0^4/U^4)} + 
\sqrt{ 4 \pi g N  } d \Omega_5^2  \right\} ~.
\cr
U_0^4 =&  { 2^7 \over 3} \pi^4  g^2  \mu
}}
We see that $U_0$ remains  finite when we take
 the  $\alpha'\to 0$ limit.  The
situation is similar to that encountered in \msfive .
Naively the whole metric is becoming of zero size since 
we have a power of $\alpha'$ in front of the metric, and 
we might incorrectly conclude that we should only consider the
zero modes of all fields. 
However, energies that are finite from the point of view of the
gauge theory, lead to proper energies (measured with respect to proper
time)
that remain finite is 
in $\alpha'$ units (or Planck units, since $g$ is fixed). 
More concretely,  an excitation that has  energy $\omega$ (fixed in the
limit)
 from 
the point of view of the gauge theory, will have 
proper energy $ E_{proper} = { 1 \over \sqrt{\alpha'} }
{ \omega  (g N 4\pi)^{1/4}\over U} $. 
This also means that the corresponding proper wavelengths 
remain fixed. 
In other words, the spacetime action on this background  has the form 
$ S \sim { 1\over \alpha'^4} \int d^{10 }x \sqrt{G} R +\cdots $, 
so  we can cancel   the factor of $\alpha'$ in the metric and 
the Newton constant, leaving a theory with a finite Planck length in the
limit. Therefore
 we should consider fields that propagate on the $AdS$ background.
Since  the Hawking temperature is finite, 
there is a flux of energy from the black hole to the $AdS$ spacetime. 
%
Since ${\cal N} = 4$ d=4 $U(N)$ SYM is a unitary theory we conclude that,
for large $N$, 
{\it it includes in its Hilbert space the states of type IIB supergravity 
on 
$(AdS_5 \times S_5)_N$}, where subscript  indicates the fact that the 
``radii''  in Planck units are proportional to $N^{1/4}$.
In 
 particular the theory  contains  gravitons 
 propagating on $(AdS_5 \times S_5)_N$.
When we consider supergravity on  $AdS_5 \times S_5$, we are faced with 
global issues like the presence of a horizon and the
 boundary conditions  at infinity. It is interesting to note 
 that the solution is nonsingular \ght . 
 The gauge 
theory should provide us with a specific choice of boundary conditions.
It would be interesting to determine them. 

We have  started with a quantum theory and we have seen that it includes
gravity so it is natural to think that this correspondence goes beyond
the supergravity approximation. 
We are led to the conjecture 
that
{\it Type IIB string theory on  $(AdS_5 \times S^5)_N$
 plus some appropriate
boundary conditions (and possibly also some boundary degrees of freedom)
 is
dual to {\cal N} =4 d=3+1 U(N) super-Yang-Mills.}
 The SYM coupling is given
by the (complex) IIB string coupling, more precisely $ { 1 \over g^2_{YM}} +
i {\theta \over 8 \pi^2}  = {1 \over  2\pi } ( { 1 \over g} 
+ i {\chi \over 2\pi})$ 
where $\chi$ is the value of the RR scalar.

The supersymmetry group of $AdS_5 \times S^5$, is known to be 
the same as the superconformal group in 3+1 spacetime dimensions 
\refs{\sohnius},
so the supersymmetries of both theories are the same. 
This is a new form of ``duality'': a large $N$ field theory is related to a 
 string theory on some background, notice that the correspondence is
non-perturbative in $g$ and the $SL(2,Z)$ symmetry of type IIB would 
follow as a consequence of the $SL(2,Z)$ symmetry of SYM\foot{
This is similar in spirit to \sltwo\ but here $N$ is not interpreted 
as momentum.}. 
It is also a strong-weak coupling correspondence in the following sense.
When the effective coupling $gN$ becomes large we cannot trust 
perturbative calculations in the Yang-Mills theory
 but we can trust calculations
in supergravity on $(AdS_5 \times S^5)_N$. 
This is suggesting that the ${\cal N} =4$
Yang-Mills master field is  the anti-deSitter
supergravity solution (similar ideas were suggested in \dps ).
Since $N$ measures the size of the geometry in Planck units, we see
that quantum effects in $AdS_5\times S^5$ have the interpretation of 
$1/N$ effects in the gauge theory. So Hawking radiation is a $1/N$
effect. It would be interesting to understand more precisely what 
the horizon means from the gauge theory point of view. 
IIB supergravity on $AdS_5\times S^5$ was studied in \refs{\sugradiv,\gun}.

The above conjecture   becomes nontrivial for large $N$ and gives a way to 
answer some large $N$ questions  in the SYM theory.
For example, suppose that we break $U(N) \to U(N-1) \times U(1)$ by
Higgsing. This corresponds to putting a three brane 
at some point on the 5-sphere and some value of $U$, with world volume
directions  along the original four dimensions ($x_{||}$). 
We  could now  ask what  the low energy effective action 
for the light U(1) fields is.
For large $N$ \sugra\ 
 it is the action of a D3 brane in $AdS_5 \times S^5$.
More concretely,  the bosonic part of the action becomes
the Born-Infeld action on the $AdS$ background 
\eqn\probeaction{\eqalign{
S &= - { 1 \over (2 \pi)^3 g } \int d^4 x  h^{-1}
 \left[ \sqrt{- Det( \eta_{\alpha \beta} + h \partial_\alpha
U  \partial_\beta U  +  U^2 h  g_{ij}\partial_\alpha
\theta^i\partial_\beta\theta^j + 
  2 \pi \sqrt{h} F_{\alpha \beta } )} -1 \right]  
\cr
  & ~~~~~~~~~h = { 4 \pi g N \over U^4 }  ~,
}}
with $\alpha, \beta = 0,1,2,3$, $i,j = 1,..,5$; and 
   $g_{ij} $ is the metric
of the unit five-sphere. 
As any low energy action, \probeaction\ is valid when the energies are
low compared to the mass of the  massive states that we are
integrating out. In this case the mass of the massive states is
proportional
to $U$ (with no factors of $N$). 
The low energy condition translates into $\p U /U \ll U$
and $\p \theta^i << U $, etc.. So the nonlinear terms in the action 
\probeaction\
will be important only when $gN$ is large. 
It seems  that the form of this action is 
completely determined by superconformal invariance, by using the broken 
and unbroken supersymmetries, in the same sense that
the Born Infeld action in flat space is given by the full Poincare
supersymmetry \schwarz .  It would be very interesting to check this 
explicitly.
We will show this for a particular term in the action.
We set $\theta^i =
 const$
and $F = 0$, so that we only have $U$ left. Then we will show that 
the action is 
completely determined by 
broken conformal invariance. 
This can be seen as follows.
Using Lorentz invariance and scaling symmetry (dimensional analysis) 
one can show that the action must  have the form
\eqn\probegen{
S = \int d^{p+1}x U^{p+1} f( \p_\alpha U \p^\alpha U/U^4) ~,
}
where $f$ is an arbitrary function.
Now we consider  infinitesimal  special conformal transformations
\eqn\espconf{\eqalign{
\delta x^\alpha  &= \epsilon^\beta x_\beta  x^\alpha -
\epsilon^\alpha ( x^2 + { \tilde R^4 \over U^2 })/2 ~, \cr
\delta U & \equiv U'(x') -U(x)   = - \epsilon^\alpha x_\alpha U ~,
}}
where $\epsilon^\alpha$ is an infinitesimal parameter. 
For the moment $\tilde R$
 is an arbitrary constant.
We will later identify it with the ``radius'' of $AdS$, it will 
turn out that $\tilde R^4 \sim g N$. 
In the limit of small $\tilde R$  we recover the more familiar  form 
of the conformal transformations ($U$ is a weight one field).
Usually  conformal transformations do not
involve the variable $U$ in the transformations of $x$.
For constant $U$ the
extra
term in \espconf\ is 
 a translation in $x$, but we will take $U$ to
be a slowly varying function of $x$ and we will determine $\tilde R$
from other facts  that we know. 
Demanding that \probegen\ is invariant under  \espconf\ 
we find that the function $f$ in \probegen\ obeys the equation
\eqn\eqforf{
f(z) + const = 2  \left(z +{ 1 \over \tilde R^4 }\right) f'(z)
}
which is solved by $ f =  b [\sqrt{1+ \tilde R^4 z} -a]$.
Now we can determine the  constants $a,b,\tilde R$ from
supersymmetry.
We need to use three facts. The first is that there is no force
(no vacuum energy) for a constant $U$. This implies $a=1$.
The second is that the $\p U^2 $ term ($F^2$ term) 
in the $U(1)$ action is not
renormalized. The third is that the only contribution to the 
$(\p U)^4$ term (an $F^4$ term) comes from a one loop diagram \dinesei .
This determines all the coefficients to be those expected from 
\probeaction\
including the fact that $\tilde R^4 = 4 \pi g N $.
%
It seems very plausible that using all 32 supersymmetries we could
fix the action \probeaction\ completely.
This would be saying that \probeaction\
 is a  consequence of the symmetries
and thus not a prediction\foot{Notice that the action
\probeaction\
 includes a term proportional to $v^6$ similar to that calculated
in  \bbpt . Conformal symmetry 
 explains the agreement that they would have found if
they had done the calculation for 3+1  SYM as opposed to 0+1.}.  
However we can make very nontrivial predictions (though we were not
able to check them). For example,
if  we take $g$ to be small (but N large)  we can predict that
the Yang-Mills theory contains strings.
 More precisely, in the limit $g \to 0$, $g N ={\rm fixed}
\gg 1$ ('t Hooft limit) we find free strings in the spectrum, they 
are IIB strings moving in $(AdS_5 \times S^5)_{gN} $.\foot{
In fact,  Polyakov \polstring\ recently  proposed that the string theory 
describing 
bosonic Yang-Mills 
 has a  new dimension corresponding to the Liouville mode
$\varphi$,
 and
that
the metric at $\varphi =0$ is  zero due to a ``zig-zag'' symmetry.
 In our case we  see that the physical distances 
along the directions of the brane contract to zero as 
$U \to 0$. The details are different, 
 since we are considering the ${\cal N}=4$ theory. }
The sense in which these strings are present is rather subtle  since
there is no energy scale in the Yang-Mills to set their tension. 
In fact one should translate the mass of a string state from the
$AdS$ description to the Yang-Mills description. This translation 
will involve the position $U$ at which the string is sitting. This 
 sets the
scale for its mass. As an example, consider 
again the D-brane probe (Higgsed configuration) which we described
above. From the type IIB description we expect 
open strings ending on the D3 brane probe. From the point of
view of the gauge theory these open strings have energies
$E =  { U \over  (4 \pi g N)^{1/4} } \sqrt{ N_{open}} $ where $N_{open}$ is the
integer charaterizing the massive open string level. In this example
we see that $\alpha'$ disappears when we translate the energies and
is replaced by $U$, which is the energy scale of the Higgs field  that 
is breaking the symmetry.

Now we turn to the question of the physical interpretation of $U$. 
$U$ has dimensions of mass. It seems natural to interpret motion in 
$U$ as moving in energy scales, going to the IR for small $U$ and to the
UV for large U. 
For example, consider a D3 brane sitting at some position $U$. 
Due to the conformal symmetry, all physics at energy scales $\omega$ in
this
theory is the same as physics at energies $\omega' = \lambda \omega$,
with the brane sitting at $U' = \lambda U$. 
%

Now let us turn to another question. We could separate a group of D3 branes
from the point were they were all sitting originally. 
Fortunately, for the extremal case we can find a supergravity 
 solution describing this
system. All we have to do is the  replacement   
\eqn\harm{
 {N \over U^4 } \to { N-M  \over U^4} + {M \over |\vec U -\vec W|^4 } ~,
}
where $\vec W =  \vec r/\alpha'$ is the separation. It is a vector because
we have to specify a point on $S^5$ also. 
The resulting metric is 
\eqn\metric{\eqalign{
ds^2 = \alpha'&\left[ 
U^2 {1 \over\sqrt{ 4\pi g } \left( N-M + {M U^4 \over |\vec U - W  |^4 }\right)^{1/2 }}
dx_{||}^2 + \right. \cr 
&+ \left. 
\sqrt{ 4\pi g } { 1 \over U^2 }  \left( N-M + {M U^4 \over |\vec U - W  |^4
}
\right)^{1/2}  d \vec U^2  \right]~.
}}
For large $U\gg |W|$ we find basically 
the solution for $(AdS_5 \times S^5)_N$ which is interpreted as saying 
that 
for large energies we do not see the fact that the conformal symmetry was
broken, while for 
small $U\ll |W| $ we find just $(AdS_5 \times S^5)_{N-M}$, which is 
associated to the CFT of the unbroken $U(N-M)$ piece. 
Furthermore, 
if we consider the region 
 $ | \vec U - \vec W | \ll |\vec W| $ we find
 $(AdS_5 \times S^5)_M$, which is described by 
the CFT of the $U(M)$ piece.

We could in principle separate all the branes from each other. For large 
values of $U$ we would still have $(AdS_5 \times S^5)_N$, but for small
values of $U$ we would not be able to trust the supergravity solution,
but we naively get $N$ copies of  $(AdS_5 \times S^5)_1$ which should
correspond to the $U(1)^N$. 

Now we discuss the issue of compactification. 
We want to consider the YM theory compactified on a torus of radii
$R_i$, $ x_i \sim x_i + 2 \pi R_i$, 
which stay fixed as we take the $\alpha' \to 0$ limit. 
Compactifying the theory breaks conformal invariance and leaves only the
Poincare supersymmetries.
However one can still find the supergravity solutions and follow the 
above procedure, going near the horizon, etc. 
The $AdS$ piece will contain some identifications. 
So we will be able to trust the supergravity solution as long as the
physical length of these compact circles stays big in $\alpha'$ units. 
This implies that we can trust the supergravity solution
as long as we stay far from the horizon (at $U=0$) 
\eqn\bigu{
 U \gg {  (gN)^{1/4}  \over R_i  }~,
}
for all $i$. 
This is a larger bound than the naive expectation 
($1/R_i$).
If we were considering near extremal black holes we would require that
$U_0$  in \ne\ satisfies \bigu , which is, of course, the same condition 
on the temperature gotten in \bfks .

The relation of the 
 three-brane supergravity solution  and the Yang-Mills theory
has been studied in 
\refs{\entrothree,\kledtdil,\dtgrav,\dtnren} . All the calculations
have been done for near extremal D3 branes  fall into the 
category described above. In particular the absorption cross section
of the dilaton and the graviton have been shown to agree with 
what one would calculate in the YM theory \refs{\kledtdil,\dtgrav}. 
It has been shown in \dtnren\ that some of these agreements are due
to non-renormalization theorems for ${\cal N} =4 $ YM. 
The black hole entropy was compared to the {\it perturbative} YM
calculation
and it agrees up to a numerical factor \entrothree . 
This is not in disagreement with the correspondence we were suggesting, 
It is expected that large $gN$ effects change this numerical 
factor, this  problem  remains unsolved.

Finally notice that  the group $SO(2,4)\times SO(6)$ suggests a
twelve dimensional realization in a theory with two times \vafaf .

\newsec{ Other cases with $16 \to 32$ 
 supersymmetries, M5 and M2  brane theories }

Basically all that we have said for the D3 brane carries over for the 
other  conformal field theories  describing 
coincident M-theory fivebranes and M-theory twobranes.
We describe below the limits that should  be taken in each of the two
cases. 
Similar remarks can be made about the entropies \kt , and the
determination of the probe actions using superconformal invariance.
Eleven dimensional supergravity on the corresponding $AdS$ spaces
was studied in \refs{\singleton,\bduff,\nst,\ght}.

\subsec{M 5 brane}

The decoupling limit is obtained by taking the 11 dimensional Planck length
 to zero, $l_p \to 0$, keeping the worldvolume energies fixed and 
taking the separations $U^2 \equiv r/l_p^3 = {\rm fixed}$ \andyfive .
 This last condition 
ensures that the membranes stretching between fivebranes give rise to
 strings
with finite tension. 

The metric is\foot{
In our conventions the relation of the Planck length to the 11 dimensional 
Newton constant is $G_N^{11} = 16 \pi^7 l_p^9 $.}
\eqn\metricfive{\eqalign{
ds^2 &= f^{-1/3} dx_{||}^2 + f^{2/3} ( dr^2 + r^2 d\Omega_4^2 ) ~,
\cr
f  & = 1 + { \pi N l_p^3 \over r^3} ~,
}}
We also have a flux of the four-form field strength on the four-sphere (which is quantized). 
Again, in the limit we obtain
\eqn\mfn{ ds^2 = l_p^2 [  { U^2 \over (\pi N)^{1/3} }
 dx_{||}^2 + 4 (\pi N)^{2/3} { d  U^2 \over  U^2 }
+ (\pi N)^{2/3}  d \Omega_4^2 ] ~,
}
where now the ``radii''  of the sphere and the $AdS_7$ space are
$R_{sph} = R_{AdS}/2 = l_p  (\pi N)^{1/3} $. 
Again, the ``radii''  are fixed in Planck units
as we take $l_p \to 0$, and supergravity can be applied if $N \gg 1$.

Reasoning as above we conclude that this theory  contains seven
dimensional Anti-deSitter times a four-sphere, which for large $N$ looks
locally like eleven dimensional Minkowski space. 

This gives us a method to calculate properties of the large N limit 
of the six dimensional (0,2) conformal field theory \swsix . 
The superconformal group again coincides with the algebra of the 
supersymmetries preserved by $AdS_7 \times S^4$. The bosonic symmetries
are $SO(2,6)\times SO(5)$ \nahm . 
We can do brane probe calculations, thermodynamic calculations \kt , etc.

The conjecture is now that {\it the (0,2) conformal field theory is dual
to M-theory on $(AdS_7 \times S^4)_{N}$ }, the subindex indicates the 
dependence of the ``radius'' with $N$.

\subsec{ M2 brane}

We now  take the limit $l_p \to 0$ keeping  $U^{1/2} \equiv 
r/l_p^{3/2} = {\rm fixed}$. This combination 
has to remain fixed because  the scalar field describing
the motion of the twobrane has scaling dimension 1/2. 
Alternatively we could have derived this conformal field theory by taking
first the field theory limit of D2 branes in string theory  as in \refs{\jmstrings,
\seimm,\senmm},
 and then 
taking the strong coupling limit of that theory to get to the conformal 
point as in \refs{\seiberg,\lsss,\bankssei}. 
The fact that the theories obtained in this fashion are the same can 
be seen as follows. 
The D2 brane gauge theory can be obtained as
the limit $ \alpha' \to 0$, keeping $g^2_{YM} \sim g/\alpha' = 
{\rm fixed}$.
This is the same as the limit of M-theory two branes in the limit
$l_p \to 0$ with  $R_{11}/l_p^{3/2} \sim g_{YM} = {\rm fixed}$.
This is a theory where one of the Higgs fields is compact. 
Taking $R^{11} \to \infty $ we see that we get the theory of coincident
M2 branes, in which  the SO(8) R-symmetry has an obvious origin.

The metric is 
\eqn\metrictwo{\eqalign{
ds^2 &= f^{-2/3} dx_{||}^2 + f^{1/3} ( dr^2 + r^2 d\Omega_7^2 )~,
\cr
f  & = 1 + { 2^5 \pi^2 N l_p^6 \over r^6} ~,
}}
and there is a nonzero flux of the dual of the four-form field strength
on the seven-sphere. 
In the decoupling limit we obtain  $AdS_4\times S^7$, and
the supersymmetries work out correctly. The bosonic generators
are given by $SO(2,3)\times SO(8)$. In this case the ``radii'' of the
sphere and $AdS_4$ are  $R_{sph} = 2 R_{AdS} =  l_p ( 2^5 \pi^2 N)^{1/6}$.

The  entropy of the near extremal solution 
agrees with the expectation  
from dimensional analysis for a conformal theory in 2+1 dimensions \kt ,
but the $N$ dependence or the numerical coefficients are not understood.

Actually for the case of the two brane the conformal  symmetry 
 was used to determine
the $v^4$ term in 
 the probe action \bfsse , we are further saying that conformal 
invariance determines it to all orders in the velocity of the  probe. 
Furthermore the duality we have proposed 
with M-theory on $AdS_4\times S^7$   determines
the precise numerical coefficient. 

When M-theory is involved the dimensionalities of the groups are
suggestive of a thirteen dimensional realization \bars .

\newsec{Theories with $8 \to 16$ supersymmetries, the D1+D5 system}

Now we consider IIB string theory compactified on $M^4$ (where $M^4 = T^4$
 or $K3$) to six 
spacetime dimensions. As a first example 
 let us start with  a D-fivebrane with four dimensions
wrapping on $M^4$ giving a string in six dimensions. Consider a system
with $Q_5$ fivebranes and $Q_1$ D-strings, where the D-string is parallel
to the  string in six dimensions arising from the fivebrane.  
This system is described at low energies by a 1+1 dimensional (4,4) 
superconformal field 
theory. 
So we  take the limit  
\eqn\lowenlim{
\alpha' \to 0~,~~~~~~~ {r\over \alpha'} ={\rm fixed} ~,~~~~~~~
 v \equiv {V_4 \over (2 \pi)^4 \alpha'^2 } =
{\rm fixed} ~, ~~~~~g_6 =  {g\over \sqrt{v} }= {\rm fixed}
}
where $V_4$ is the volume of $M^4$. All other moduli of $M^4$ remain fixed.
This is just a low energy limit, we keep all dimensionless moduli fixed. 
As a six dimensional theory, IIB on $M^4$ 
contains strings. They transform under the U-duality group and they
carry charges given by a vector $q^I$. In general we can 
consider a configuration
where $q^2 = \eta_{IJ} q^I q^J \not = 0$ (the metric is the U-duality  
group invariant), and then take the limit \lowenlim.

This theory has a branch which we will call the Higgs branch and 
one which we call the Coulomb branch. On the Higgs branch  the 
1+1 dimensional vector multiplets have zero expectation value
and the Coulomb branch is the other one. Notice that 
 the expectation values of the vector multiplets in the
Coulomb branch remain  fixed as we take the limit $\alpha' \to 0$. 

The Higgs branch 
is a SCFT with (4,4) supersymmetry. This is the theory considered in \ascv .
The above limit includes also a piece of the  Coulomb branch, 
since we can separate the branes by a distance such that the
mass of stretched strings remains finite. 

Now we consider the  supergravity solution corresponding
to D1+D5 branes \hms
\eqn\fdbh{ \eqalign{
ds^2 &= f_1^{-1/2} f_5^{-1/2} dx_{||}^2  +  f_1^{1/2} f_5^{1/2}(
dr^2 + r^2 d \Omega_3^2 ) ~,
\cr
f_1 &= \left( 1 + { g \alpha' Q_1 \over v r^2}
\right) ~,~~~~~~~~~~
f_5 = \left( 1 + {g \alpha' Q_5 \over  r^2} \right)~,
}}
where 
$dx_{||}^2 = -dt^2 + dx^2$ and  $x$ is the coordinate along the
D-string. Some of the  moduli of $M^4$ vary over the solution 
and attain a fixed value at the horizon which depends only on
the charges and some others are constant throughout the solution.
The three-form RR-field strength is also nonzero.

%

In the decoupling limit \lowenlim\
 we can neglect the 1's in $f_i$ in \fdbh\
and  the metric becomes 
\eqn\fdbh{ 
ds^2 = \alpha'\left[
 { U^2 \over g_6 \sqrt{ Q_1Q_5 }} dx_{||}^2 + g _6\sqrt{ Q_1 Q_5}
 { dU^2 \over U^2 }
+ g_6 \sqrt{ Q_1 Q_5 }
 d\Omega_3^2 
\right]~.
}
The compact manifold $M^4(Q)$ that results in the limit is determined as follows.
Some  of its 
moduli 
are at their fixed point value which depends only on the
charges and not on the asymptotic value of those moduli 
at infinity (the notation $M^4(Q)$ 
indicates the charge dependence of the moduli)\ferrarafm \foot{
The fixed values of the moduli are determined by the condition
that they  minimize the
tension of the corresponding string (carrying charges $q^I$) 
in six dimensions 
\ferrarafm . This is parallel to the discussion 
in four dimensions \fixedmod .}. 
The other moduli, that were constant in the black hole solution, have 
their original  values. 
For example, the volume of $M^4$ 
has the  fixed point value $v_{fixed}  = 
Q_1/Q_5$, while the six dimensional string coupling $g_6$ has the 
original value.  
Notice that there is an overall factor of $\alpha'$ in \fdbh\  which can  be 
removed by canceling it with the factor of $\alpha'$ in the 
Newton constant as explained above.  
We can trust the supergravity solution if 
 $Q_1, Q_5$  are  large,  $g_6  Q_i \gg 1$. 
Notice that we are talking about a six dimensional 
supergravity solution since the volume of $M^4$ is a constant
in Planck units (we keep the $Q_1/Q_5$ ratio fixed). 
The metric \fdbh\ describes three dimensional $AdS_3$ times a 3-sphere.
The supersymmetries work out correctly, starting from the 8 Poincare 
supersymmetries 
we enhance then to 16 supersymmetries. The bosonic component
is  $SO(2,2) \times SO(4)$. In conformal field theory language
$SO(2,2)$ is just the $SL(2,R) \times  SL(2,R)$ 
part of the conformal group and
SO(4) $\sim SU(2)_L \times SU(2)_R $ are the R-symmetries of the CFT \spin .

So the conjecture is that {\it  the 1+1 dimensional CFT describing
the Higgs branch of the D1+D5 system on $M^4$  
is dual to 
type IIB string theory on $(AdS_3\times S^3)_{Q_1Q_5}\times M^4(Q)$ }.
The subscript indicates that the radius of the three
sphere is $R^2_{sph} = \alpha' { g_6 \sqrt{Q_1 Q_5} }$. 
 The compact fourmanifold $M^4(Q)$ is at some
particular point in moduli space determined as follows.
The various moduli of $M^4$ are divided as tensors  and hypers according to
the (4,4) supersymmetry on the brane.
Each hypermultiplet contains four moduli and each tensor contains a 
modulus and an anti-self-dual $B$-field. (There
are five tensors of this type for $T^4$ and 21 for $K3$). 
 The scalars in the tensors  have fixed point 
values at the horizon of the black hole, and those values are the
ones entering in the definition of $M^4(Q)$ ($Q$ indicates the dependence
on the charges). The hypers will have the
same expectation value everywhere. 
It is necessary for this conjecture to work that the 1+1 dimensional 
(4,4) theory is indendent of the tensor moduli appearing in its original 
definition as a limit of the brane configurations, since $M^4(Q)$ does not
depend on those moduli. 
A non renomalization theorem like \refs{\jmmlow, \seidia} would explain this. 
We also need that the Higgs branch decouples from the Coulomb branch as
in \refs{\abkss,\edhiggs}.

Finite temperature configurations in the 1+1 conformal field 
theory can be considered. They  correspond to near extremal black holes
in $AdS_3$. The metric is the same as that of the BTZ 2+1 dimensional 
black hole \btz ,
except  that the angle of the BTZ description is not periodic. This angle
 corresponds
to the spatial direction $x$ of the 1+1 dimensional CFT and 
it becomes  periodic if we  compactify the 
theory\foot{
I thank G. Horowitz for many discussions on this correspondence and for
pointing out ref. \koreano\ to me. Some
 of the  remarks the remarks below arose in conversations with him.} 
 \refs{\koreano,\kostas,\sfkostas} \foot{
The ideas   in \refs{\koreano,\kostas,\sfkostas} could be used to 
show the relation between the $AdS$ region and black holes
in  M-theory on a light like circle. However the statement
in \refs{\koreano,\kostas,\sfkostas} that the  
 $AdS_3 \times S^3$ 
 spacetime is U-dual to the full black hole solution (which is 
asymptotic to  Minkowski space) should be taken with caution  
because in 
those cases the spacetime has identifications on circles that
are becoming null. This  changes dramatically  the physics.
 For  examples  of these changes  see \refs{\seimm,\polch}. }.
All calculations  done for the 1D+5D system \refs{\ascv,\ghas,\greycalc}
are  evidence for this conjecture. In all these cases \greycalc\ the
nontrivial part of the greybody factors comes from the $AdS$ part of 
the spacetime. Indeed, it was noticed in \irish\ that
the greybody factors for the BTZ black hole were the same as the
ones for the five-dimensional black hole in the dilute gas approximation.
The dilute gas condition $r_0, r_n \ll r_1 r_5 $ \ghas\  is 
automatically satisfied
in the limit \lowenlim\ for finite temperature configurations
 (and finite chemical potential for
the
momentum along $\hat x$).
 It was also noticed that the 
equations have an $SL(2,R) \times SL(2,R)$ symmetry \finn , these 
 are  the
isometries   of $AdS_3$, and part of the
 conformal symmetry of the 1+1 dimensional
field
theory.  
It would be interesting to understand what  is the gravitational 
counterpart of the full conformal symmetry group in 1+1 dimensions.

%





\newsec{Theories with $ 4\to 8$  supersymmetries}

The theories of this type will be related to  black strings
in five dimensions and \RN\ black holes in four dimensions. 
This part will be more sketchy, since there are several details 
of the conformal field 
theories involved which I do not completely understand, most notably 
 the dependence on the various moduli of the compactification.

\subsec{Black string in five dimensions}

One can think about this case as arising from M-theory on $M^6$ where
$M^6$ is a CY manifold,  $K3\times T^2$ or $T^6$. 
We wrap fivebranes on a four-cycle $P_4 = p^A \alpha_A$ 
in $M^6$ with nonzero triple 
self intersection number, see \msw .
 We are left with a one dimensional object 
in five spacetime dimensions. 
Now we take the following limit
\eqn\limittf{
l_p \to 0~~~~~~~~~~~ (2\pi)^6 v \equiv V_6/l_p^6 = 
{\rm fixed }~~~~~~~ U^2 \equiv r/l_p^3 = {\rm fixed}~, 
}
where $l_p$ is the eleven dimensional Planck length. 
In this limit the theory will reduce to a conformal field theory 
in two dimensions. It is   a (0,4) CFT and it was discussed in 
some detail in a  region of the moduli space in \msw .
More generally we should think that the five dimensional theory has
some strings characterized by  charges $p^A$, forming a multiplet
of the U-duality group and we are taking a configuration where
the triple self intersection number $ p^3$ is nonzero (in the case 
$M^6=T^6$,  $p^3 \equiv D \equiv D_{ABC} p^A p^B p^C $ 
is the cubic $E_6$ invariant). 

We now  take the corresponding limit of the black hole solution. 
We will just present the near horizon geometry, obtained after
taking the limit. Near the horizon all the vector 
moduli  are at their fixed point values \fmf . So the near horizon 
geometry can be calculated by considering the solution with
constant moduli. We get 
\eqn\metf{
ds^2 = l_p^2 \left[ { U^2 v^{1/3} \over D^{1/3}} (-dt^2 + dx^2) + 
 { D^{2/3}\over v^{2/3} }  \left(4 { dU^2 \over U^2} + 
 d\Omega_2^2 \right) 
\right] ~.
}
In this limit 
 $M^6$ has its  vector moduli equal to their fixed point values
which depend only on the charge
while its  hyper  moduli are  what
they were at infinity.
  The overall size  of $M^6$ in Planck units
is a hypermultiplet, so
it remains constant as we take the limit \limittf .
We get a product of three dimensional $AdS_3$ spacetime  with a two-sphere,
$AdS_3\times S^2$.
Defining the five dimensional Planck length by 
$l_{5p}^3 = l_p^3/v$ we find that 
the ``radii''  of the two sphere and the $AdS_3 $ are
$R_{sph} = R_{AdS}/2 = l_{5p} D^{1/3}$.
%
In this case the superconformal group  contains as  a bosonic
subgroup  $SO(2,2)\times SO(3)$. So the R-symmetries are just $SU(2)_R$,
associated to the 4 rightmoving supersymmetries.

In this case we conjecture that 
this (0,4) conformal field theory 
%
is dual, 
for large $p^A$, to   M-theory on $
 AdS_3\times  S^2 \times M^6_{p}$.  The hypermultiplet  moduli  
of $M^6_p$ are  the same as the
ones entering the definition of the (0,4) theory. The vector  moduli 
depend only on the charges and their values are those that 
 the black string has at the horizon. 
A necessary condition 
 for this conjecture to work is that the (0,4) theory  
 should be independent of
the original values of the vector
moduli (at least for large $p$). It is not clear to me whether this is
true.

Using this conjecture 
 we would get for large $N$ a compactification 
of $M$ theory which has five extended dimensions.

\subsec{Extremal 3+1   dimensional \RN }

This section is  more sketchy and contains an unresolved puzzle,
so the reader will not miss much if he skips it.

We start with IIB string theory compactified on 
 $M^6$, where $M^6$ is a Calabi-Yau manifold
 or $K3\times T^2$
or $T^6$. 
We consider a configuration of $D3$ branes that leads to a black hole with
nonzero horizon area. 
Consider the limit
\eqn\limfourd{
\alpha' \to 0 ~~~~~~~~~ 
(2\pi)^6 v \equiv {V_6 \over \alpha'^3 } = {\rm fixed} ~~~~~~~~~
U \equiv  {r \over
\alpha'}
= {\rm fixed}~.
}
The string coupling is arbitrary. 
In this limit the system reduces to quantum mechanics on the moduli 
space of the three-brane configuration.

Taking the limit \limfourd\ of the supergravity solution we find 
\eqn\twod{
ds^2 = \alpha'\left[
{  U^2 \over g^2_4 N^2} dt^2 + {g^2_4 N^2 } { d U^2 \over U^2} + 
{ g^2_4 N^2 } d\Omega_2^2
 \right] }
where $N$ is proportional to the number of D3 branes. 
We find a two dimensional $AdS_2$ space times a two-sphere, both
with the same radius $R =  l_{4p} N$,  where $l_{4p}^2 = g^2 \alpha'/v$.
The bosonic symmetries of $AdS_2\times S^2$
 are $SO(2,1)\times SO(3)$.
This superconformal symmetry seems related to the
symmetries of the {\it chiral} conformal field theory 
that was proposed in \msangular\ to describe the \RN\ black holes.
Here we find a puzzle, 
since  in the  limit \limfourd\ we got
a quantum mechanical system and not a 1+1 dimensional 
conformal field theory. 
In the  limit \limfourd\   the energy gap (mentioned in 
\refs{\mss,\msangular})  becomes
very large\foot{
I thank A. Strominger for pointing this out to me.}.
 So it looks like taking a large $N$ limit at the same
time will  be crucial in this case.
 These  problems might
be related to the large ground state entropy of the system.  

If this is  understood  it might lead to  a proposal  
 for a non perturbative
definition of M/string  theory (as a large $N$ limit) when there are four 
non-compact dimensions.

It is interesting to consider the motion of probes on the
$AdS_2$ background. This corresponds to going into
the ``Coulomb" branch of the quantum mechanics. 
Dimensional analysis says that the action has the 
form \probegen\ with $p=0$.
Expanding $f$ to first order we find 
$S \sim \int dt {  \dot U^2 \over U^3} \sim \int dt v^2/r^3 $,
which is the dependence on $r$ that we expect
from supergravity when we are close to the horizon.
A similar analysis for \RN\ black holes in five dimensions
would give a term proportional to $1/r^4$ \dps .
It will be interesting  to check the coefficient 
(note that this is the {\it only } term allowed by
the symmetries, as opposed to \dps ).

\newsec{ Discussion, relation to  matrix theory}

By
deriving various field theories from string 
theory 
 and considering their large $N$ limit we have shown that 
they   contain in their Hilbert space excitations
describing supergravity on various spacetimes. 
We further conjectured that  the field  theories  are dual to the full quantum
M/string theory on   various spacetimes. 
In principle, we can use  this duality to give a  definition
of M/string theory on flat spacetime as (a region of)  the large $N$ limit of the
field theories. 
Notice that this is a non-perturbative proposal  for defining such 
theories, since the corresponding field theories can, {\it in principle}, 
be defined non-perturbatively. We are only scratching the surface and
there are many things to be worked out.
In \bfss\ it has been proposed that the large $N$ limit of 
D0 brane quantum mechanics would describe eleven dimensional M-theory.
%
The large $N$ limits discussed above, also provide  a definition 
of M-theory. An obvious  difference with the matrix model of \bfss\ is 
that here $N$ is not interpreted as  the momentum along a compact 
 direction. In our 
 case,  $N$ is related to the curvature and the size 
of the space where the theory is defined. In both cases, in the large 
$N$ limit we expect to  get flat, non-compact spaces. 
The  matrix model \bfss\ gives us a prescription 
 to build asymptotic states,
we have not shown here how to construct 
graviton states, and this is a very interesting problem. 
On the other hand, with the present proposal it is more clear 
 that 
we recover supergravity in the large $N$ limit. 

This approach leads to proposals involving five (and
maybe in some future four)
 non-compact
dimensions. The five dimensional proposal involves considering 
the 1+1 dimensional field theory associated to a black string 
in five dimensions.
These theories need to be studied in 
much more detail than we have done here.

It seems that this  correspondence between the large 
$N$ limit of field theories and supergravity can be extended to 
non-conformal field theories. 
An example was considered in \msfive , where the theory of NS fivebranes
was studied in the $g \to 0$ limit. 
A  natural interpretation for the throat region is that it is a region
in the Hilbert space of a six dimensional ``string" theory\foot{
This possibility was also raised by  \seiprivate  , though it is a bit disturbing
to find a constant energy flux to the UV (that is how we are interpreting the radial dimension).}.  
And the fact 
that contains gravity in the large N limit is just a common  feature
of the large $N$ limit of various field theories. 
The large $N$ master field seems to be the anti-deSitter supergravity 
solutions \dps .

When we study non extremal 
black holes in $AdS$ spacetimes we are no longer
restricted to low energies, as we were in the discussion in higher
dimensions \refs{\jmmlow,\greycalc}. 
The restriction came from matching the $AdS$ region to the 
Minkowski region.  So the five dimensional results  
\refs{\ghas,\greycalc} can be used to 
describe arbitrary non-extremal black holes in three dimensional 
Anti-deSitter spacetimes. This might lead us to understand better where
the degrees of freedom of black holes really are, as well as the 
meaning of the region behind the horizon.
The question of the boundary conditions is very interesting and
the conformal field theories should provide us with some
definite boundary conditions and will probably explain us how
to interpret physically  spacetimes with horizons. 
It would be interesting to find the connection
with the description of 2+1 dimensional black holes 
proposed by Carlip \carlip . 

In \refs{\singleton,\ckvp}
super-singleton representations of $AdS$ were studied
and it was proposed that they would describe the dynamics of a brane
``at the end of the world''. It was also found that in maximally
supersymmetric
cases it reduces to a free field \singleton .
It is tempting therefore to identify
the singleton  with the center of mass degree of freedom of the branes
\refs{\ggpt,\ckvp}. A recent paper suggested that super-singletons
would describe all the dynamics of $AdS$ \sfkostas . The claim
of the present paper is that all the dynamics of $AdS$ reduces to 
previously known
conformal field theories.

It seems natural  to study conformal field theories in Euclidean space  
and relate them to deSitter spacetimes.
 
Also it would be nice if these results could be extended to 
four-dimensional gauge theories with less supersymmetry.

{\bf Acknowledgments}

I thank specially G. Horowitz and A. Strominger for many discussions.
I also thank R. Gopakumar, 
 R. Kallosh, A. Polyakov,  C. Vafa and E. Witten for discussions 
at various stages in this project. 
My apologies to everybody   I did not cite in the previous version of
this paper. I thank the authors of \kal\ for pointing out a sign error.

This work was supported in part by 
DOE grant
DE-FG02-96ER40559.

\newsec{Appendix}

$D=p+2$-dimensional anti-deSitter spacetimes can be obtained by taking 
the hyperboloid 
\eqn\hyperb{ - X^2_{-1}  - X_0^2 + X_1^2 + \cdots + X_p^2 + X_{p+1}^2 = 
- R^2 ~,
}
embedded in a flat D+1 dimensional spacetime with the metric 
$ \eta = Diag( -1,-1,1,\cdots,1)$. 
We will call $R$ the ``radius'' of $AdS$ spacetime. 
The symmetry group 
$SO(2,D-1)= SO(2,p+1)$ is obvious in this description. 
In order to make contact with the previously presented form  of the metric 
let us define the coordinates
\eqn\newcoord{\eqalign{
 U &= ( X_{-1} + X_{p+1}) \cr
x_\alpha &= {X_\alpha R \over   U }~~~~~~~~~~~\alpha = 0,1,\cdots,p \cr
V  & =  (X_{-1} - X_{p+1})  = { x^2  U \over R^2  } 
+ {R^2 \over U }~ .
}}
The induced metric on the hyperboloid \hyperb\ becomes
\eqn\indmet{
 ds^2 =  {  U^2 \over R^2 } dx^2 + R^2 { d  U^2 \over  U^2}~.
}
This is  the form of the metric used in the text.
We could also define $\tilde U = U/R^2$ so  that metric \indmet\
has  an overall factor of $R^2$, making it clear that $R$ is the
overall scale of the metric. 
The region outside the horizon corresponds to $U>0$, which is 
only a part of \hyperb . It would be interesting to understand what
the other regions in the $AdS$ spacetime correspond to. For 
further discussion see \hawellis .


\listrefs

\bye